\documentclass[12pt]{iopart}
\usepackage{iopams} 
\usepackage{color}
\usepackage{cite}
\usepackage{graphicx} 
\usepackage{subfigure} 
\usepackage{tikz} 
\usepackage{epstopdf} 
\begin{document}

\title[Transport coefficients for a confined Brownian ratchet]
{Transport coefficients for a confined Brownian ratchet operating between two heat reservoirs}

\author{A~Ryabov$^{1  \dag}$, V~Holubec$^{1}$, M~H~Yaghoubi$^{2,3}$, M~Varga$^{1}$, M~E~Foulaadvand$^{2,4}$, and P~Chvosta$^{1}$}

\address{$^{1}$ Charles University in Prague, Faculty of Mathematics and Physics, Department of Macromolecular Physics, V Hole{\v s}ovi{\v c}k{\' a}ch 2, 180~00~Praha, Czech Republic}
\address{$^{2}$ Department of Physics, University of Zanjan, P.O.~Box~45196-311, Zanjan, Iran}
\address{$^{3}$ Complexity Science Group, Department of Physics \& Astronomy, University of Calgary, Calgary, Alberta~T2N~1N4, Canada}
\address{$^{4}$ School of Nano-Science, Institute for Research in Fundamental Sciences (IPM), P.O.~Box~19395-5531, Tehran, Iran}

\ead{$^{\dag}$rjabov.a@gmail.com}


\begin{abstract}
We discuss two-dimensional diffusion of a Brownian particle confined to a periodic asymmetric channel with soft walls modeled by a parabolic potential. In the channel, the particle experiences different thermal noise intensities, or temperatures, in the transversal and longitudinal directions.  The model is inspired by the famous Feynman's ratchet and pawl. Although the standard Fick-Jacobs approximation predicts correctly the effective diffusion coefficient, it absolutely fails to capture the ratchet effect. Deriving a correction, which breaks the local detailed balance with the transversal noise source, we obtain a correct mean velocity of the particle and a stationary probability density in the potential unit cell. The derived results are exact for small channel width. Yet, we check by exact numerical calculation that they qualitatively describe the ratchet effect observed for an arbitrary width of the channel.
\end{abstract}

\maketitle 
\section{Introduction} 

Microscopic devices capable of extracting work from thermal fluctuations have attracted considerable interest since origins of statistical mechanics. Perhaps the best known of these Maxwell demons \cite{MaxDemon} is the ratchet model due to Smoluchowski \cite{Smoluchowski1912} and Feynman \cite{Feynman}. The model consists of 1) a wheel with asymmetric teeth (see Fig.~$2.1.$~in Ref.~\cite{Reimann}, or Fig.~\ref{fig:model} below), randomly rotating due to perpetual bombardment by molecules of a surrounding fluid; and 2) a pawl which is aimed to rectify the rotatory Brownian motion of the wheel. Apparently, such a device is capable to transform erratic thermal fluctuations into systematic motion with the potential to perform work on an external load attached to the wheel, and indeed, when the wheel and the pawl are coupled to heat reservoirs {\em at different temperatures}, the wheel starts to rotate in one direction \cite{Feynman}.

Though originally introduced by Feynman as a pedagogical tool, the two-temperature ratchet has gained  significant attention in recent years. The model belongs to the class of non-equilibrium steady-state stochastic heat engines and it is interesting from both the dynamic \cite{Reimann} and the thermodynamic \cite{SekimotoBook, Seifert2012, OonoPaniconi1998, HatanoSasa2001, Komatsu2010} points of view.  Since the original Feynman's mechanical setting is too complex to be analyzed from the first principles (see, however,~\cite{VanDenBroeck2004, 2016arXiv160509142F}), several {\em qualitatively similar} stochastic models emerged \cite{Sekimoto1997, Magnasco1998, Hondou1998, JarzynskiWojcik2004, KomatsuPRE2006, Gomez-Marin2006, RoeckMaes2007, KomatsuPRL2008, Ueda2012, JarzynskiMazonka1999}.
A couple of them are exactly solvable \cite{JarzynskiMazonka1999,  Visco2006, Fogedby2011, Dotsenko2013, Grosberg2015}, other  were analyzed numerically \cite{SekimotoBook, Tumlin2016}. 
The models were studied regarding their reversibility \cite{Hondou1998} and from a general point of view of symmetry breaking \cite{RoeckMaes2007}. Fluctuation theorems for such models were derived in Refs.~\cite{JarzynskiWojcik2004, Gomez-Marin2006} and in Refs.~\cite{KomatsuPRL2008, NakagawaKomatsuEPL2006} general results regarding steady state were derived in the linear-response regime. 
Experimentally, stochastic heat flow in a two-temperature microscopic system was studied with optical tweezers \cite{CilibertoPRL2013,Berut2016}.

In the present article we propose and solve a two-dimensional Brownian model of the Feynman's ratchet (see Fig.~\ref{fig:model} and Sec.~\ref{sec:model}). The model is based on the Brownian motion in the confining asymmetric channel subjected to thermal noises with different intensities (temperatures) along the transversal and longitudinal directions.  The model is solved by a generalization of the so-called Fick-Jacobs approximation \cite{Zwanzig1992, RegueraRubi, CPHC:CPHC200800526}. The presented expansion allows us to derive principal characteristics of the ratchet -- the stationary PDF in the unit cell of the potential, the mean velocity of the particle, and the effective diffusion coefficient as the functions of the two temperatures.

The discussed model offers a rare possibility to study analytically a rather non-trivial physics of a two-dimensional far-from-equilibrium stochastic system. At the same time it provides further physical insight into the Fick-Jacobs theory. The latter was originally developed to characterize impact of entropic forces on diffusion in narrow corrugated channels \cite{Zwanzig1992, RegueraRubi, CPHC:CPHC200800526} and during recent years it has become a subject of an intensive testing and extensions \cite{KalinayPercus2005, KalinayPercusPRE2005, KalinayPercusPRE2006, BuradaPTRA2009, BuradaSchmid2009, Drazer2009, Leo2010JCP, LeoBerezh2010JCP, KalinaySoft2011, KalinayPRE2011, MartensPRE2011, MartensChaos2011, PinedaLeoJCP2011, Dagdug2012, DagdugPinedaJCP2012, Martens2012Communication, KalinayJCP2013, LeoJCP2013, Martens2013, MartensPRL2013, Metzler2014, KalinaySinusDrivingPRE2014, Leo2014PhysicaA, Leo2014PRE,  Drazer2015, DasRayPRE2015, Leo2015JCP, Wang2016, Leo2016CW, Malgaretti2016, Malgaretti2016Polymer, KalinayPRE2016}. 
Various Brownian ratchets have already been studied with the aid of the Fick-Jacobs approximation including flashing and rocking ratchets  \cite{ConfinedRubi2012, Makhnovskii2012, ConfinedRubi2013} and ratchets driven by a temperature gradient  \cite{ConfinedRubi2013}. Actual applications of the ratchet effect in  corrugated channels for separation of particles according to their sizes was investigated both theoretically~\cite{SlaterElectrolytes97, RegueraPRL2012} and experimentally~\cite{PhysRevLett.88.168301, C2LC21089D}. 
In the present work, we develop a perturbation expansion, which generalizes the Fick-Jacobs theory to the case of diffusion in the corrugated channel with soft walls (i.e., the particle is confined by the potential instead of the hard walls which are commonly considered in the literature) driven out of equilibrium by two thermal baths. In the previously studied models, the behavior of the mean particle velocity was always captured by the lowest order approximation, i.e., by the standard Fick-Jacobs theory. In our case, the fact that the particle transport is driven solely by temperatures of the two heat reservoirs, introduces substantial complications. 
The ratchet effect is absent in the lowest-order approximation (mean particle velocity is zero) and it appears only in higher-order terms. This behavior is a direct consequence of the assumptions behind the Fick-Jacobs theory (see the last two paragraphs of Sec.~\ref{sec:lowest-order}).  

In Sec.~\ref{sec:model} we define the model, describe its basic physics and sum up the derived results. Sec.~\ref{sec:FP} comprises the outline of the used perturbation theory for the Fokker-Planck equation. In Sec.~\ref{sec:v} the derived mean velocity is discussed on physical grounds and tested against numerics. Sec.~\ref{sec:DiffusionCoeff} comprises the discussion and verification of the derived diffusion coefficient. The numerical approach  is explained in the~\ref{App:numerics}.

\begin{figure}[t!]
\centering
\qquad
\subfigure{
\begin{tikzpicture}[scale=0.46,
      trans/.style={thick,shorten >=2pt,shorten <=2pt,>=stealth},
]
\draw [black,thick,<->,domain=30:-60] plot ({1*cos(\x)}, {1*sin(\x)});
\draw (0,0) [thick] circle (0.4cm);
\draw[trans,<->] (2.5,-4.5) -- (4.5,-4.5) node[midway,midway] {};
\draw [fill=black] (4.5,0.3) rectangle (2.5,0.1);
\draw [fill=black] (3.55,0.1) rectangle (3.45,-6.0);
\draw (1.5,-7.0) rectangle (5.5,-3.5);
\node[draw=none] at (4.5,-5.5) {$T_y$};
\draw (-3,-2.6) rectangle (10.0,2.6);
\node[draw=none] at (3.5,2) {$T_x$};
\draw[thick]
\foreach \i in {1,2,...,10} {%
   [rotate=(\i-1)*36]  (0:2)  arc (5.0:12:2) -- (20:2.3)  arc
(18:30:2.4) --  (36:2)
 };
\begin{scope}[xshift=7.0cm,rotate=180/8]
\draw [black,thick,<->,domain=7.5:-82.5] plot ({1*cos(\x)}, {1*sin(\x)});
\draw (0,0) [thick] circle (0.4cm);
\draw[thick]
\foreach \i in {1,2,...,10} {%
   [rotate=(\i-1)*36]  (0:2)  arc (5.0:12:2) -- (20:2.3)  arc
(18:30:2.4) --  (36:2)
 };\end{scope}
\draw[fill=white,thick] (1.1,0.9) rectangle (8.1,1.3);
\draw [fill=white,thick] (1.1,1.1) circle (0.2cm);
\draw [fill=white,thick] (8.1,1.1) circle (0.2cm);
\end{tikzpicture}
}
\qquad 
\subfigure{
\includegraphics[width=0.42\linewidth]{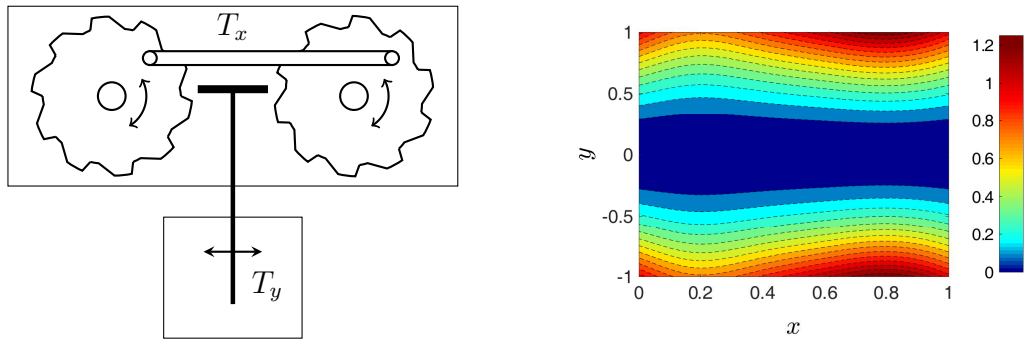}
}
\caption{Left: Schematic of the mechanical ratchet corresponding to our model. The synchronized wheels randomly rotate due to impacts of molecules at the temperature~$T_x$. The pawl, placed between the wheels, moves randomly due to the coupling with the heat bath at the temperature~$T_y$.  
Right: Equipotentials within one period of the parabolic potential (\ref{potentialGEN}) with the asymmetric spring constant given in Eq.~(\ref{SpringConst}), $\varepsilon=1$. The two-temperature Brownian motion in the potential models the  mechanical ratchet.}
\label{fig:model} 
\end{figure}

\section{The model}
\label{sec:model}
The two-temperature mechanical ratchet  corresponding to our model is depicted in the left panel of Fig.~\ref{fig:model}. 
We describe the dynamics of the mechanical ratchet by a two-temperature overdamped Brownian motion in a two-dimensional channel with asymmetric walls. In the long-time limit, the overdamped dynamics is expected to yield  qualitatively similar results as the more general underdamped Brownian motion \cite{SekimotoBook, NakagawaKomatsuJPSJ2005, NakagawaKomatsu2005}, which, however, appears  to be  analytically intractable. The two coordinate axes are chosen as follows. The coordinate $x$ is associated with the angle of rotation of the wheels in the mechanical ratchet. In the Brownian model, the $x$ axis points along the channel central line. The confining potential, which models the repulsion of the pawl from the asymmetric teeth, is periodic in $x$ direction. One period of the potential is plotted in the right panel of Fig.~\ref{fig:model}.  In the transversal direction $y$ (position of the pawl) the potential increases without bounds thus confining the particle to diffuse along $x$. 
We model the channel  walls by the parabolic potential  
\begin{equation}
\label{potentialGEN} 
U(x,y)= \frac{k(x)}{ 2\varepsilon^{2} } y^{2}, \qquad k(x) = k(x+1),
\end{equation} 
where the $1$-periodic ``spring constant'' $k(x)/\varepsilon^{2}$ controls the width of the channel, which becomes very narrow for small $\varepsilon$. 
The asymmetry of ratchets teeth in the mechanical model is reflected in the asymmetry of $k(x)$. For the sake of illustrations we choose the usual \cite{Reimann} asymmetric function
\begin{equation}
\label{SpringConst} 
k(x) = 2-\frac{\sin(2\pi x)}{2} -\frac{ \sin(4 \pi x)}{12}.
\end{equation} 
 Equipotentials of (\ref{potentialGEN}) with (\ref{SpringConst}) are illustrated in the right panel of Fig.~\ref{fig:model}.

\begin{figure}[t!]
\begin{center} 
\includegraphics[width=0.85\linewidth]{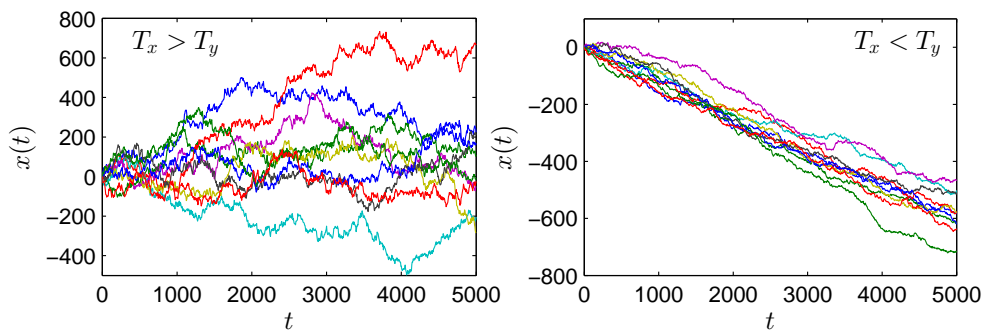}
\caption{Trajectories of the two-temperature Brownian motion in the periodic potential~(\ref{potentialGEN}) for $T_x > T_y$ (the left panel; $T_x = 10$, $T_y = 1$) and $T_x < T_y$ (the right panel; $T_x = 1$, $T_y = 10$). Other parameters are $\varepsilon^{2}=0.01$ and $dt = 5 \times 10^{-6}$, where $dt$ is the time step used in the numerical integration of the Langevin equations (\ref{Langevin}) by the Euler-Maruyama method \cite{bookKloeden}. }
\label{fig:trajectories}
\end{center}
\end{figure} 

In the channel, the particle experiences different noise intensities, or temperatures, in the longitudinal and the transversal directions. The transversal temperature $T_y$ is associated with the temperature of the pawl, the heat bath at the temperature $T_x$ causes stochastic rotation of the wheels. 
In the renormalized units, where $k_{\rm B}=1$ and the mobilities in both directions are equal to one, the Langevin equations for the particle coordinates $x(t)$, $y(t)$, read 
\begin{equation} 
\label{Langevin}
\frac{{ d} x}{{ d}t} = -   \frac{k'(x) }{2 \varepsilon^{2}}y^{2}  + \sqrt{2 T_x }\, \xi_{x}(t), \qquad 
\frac{{ d} y}{{ d}t} = - \frac{k(x)}{\varepsilon^{2}} y + \sqrt{2 T_y }\, \xi_{y}(t),  
\end{equation}
where $\xi_{i}(t)$ is the delta-correlated Gaussian white noise:  $\left<\xi_{i}(t)\xi_{j}(t')\right> = \delta_{ij} \delta(t-t')$, $\left<\xi_{i}(t)\right> =0$, $i,j=x,y$. 
Numerical integration of the Langevin equations reveals an intriguing property of the model: when $T_x \neq T_y$,  the particle moves on average in one direction along the channel. A couple of trajectories of $x(t)$ are plotted in Fig.~\ref{fig:trajectories}. Notice that for $T_x>T_y$ the diffusive motion along $x$ is strongly fluctuating and it is impossible to decide, with the naked eye, whether there is any systematic drift or not. On the other hand, when $T_x < T_y$, the fluctuations are less pronounced and a clearly visible drift  emerges.

The main goal of the present study is to put on firm analytical grounds what we have just observed on the level of individual trajectories. We will derive the mean (longitudinal) particle velocity $v(T_x, T_y)$ and the effective (longitudinal) diffusion coefficient $D(T_x, T_y)$, 
\begin{equation} 
\label{vD_deff}
v(T_x, T_y) = \lim_{t\to \infty} \frac{\left< x(t) \right> }{t},
\qquad 
D(T_x, T_y) = \lim_{t\to \infty} \frac{\left< \left[ x(t) - \left<x(t) \right>  \right]^{2} \right> }{2 t},
\end{equation}
as the functions of the two temperatures. Our analytical approach is based on the expansion of the Fokker-Planck equation for small $\varepsilon$ (narrow channel) and it is similar in spirit to that used previously by Laachi et al.\ \cite{Laachi2007} for hard-wall channels. We find that the mean velocity is the second-order effect in the channel width, i.e., 
\begin{equation}
v(T_x, T_y) = \varepsilon^{2} v_{1}(T_{x},T_{y} ) + O(\varepsilon^{4}),
\end{equation}
where the function $ v_{1}(T_{x},T_{y} )$ is given in Eq.~(\ref{v1TxTy}) and it is depicted in Figs.~\ref{fig:vTy}~and~\ref{fig:vTx}, where it is compared to the numerical results obtained by discretization of the Fokker-Planck equation as explained in~\ref{App:numerics}. 
  
Contrary to the mean velocity, the effective longitudinal diffusion coefficient is well approximated by the lowest order term in the expansion 
\begin{equation} 
\label{Dexpanded}
D(T_x, T_y) = D_{0}(T_{x},T_{y} ) + \varepsilon^{2}  D_{1}(T_{x},T_{y} ) + O(\varepsilon^{4}).
\end{equation}
The analytical result for $D_0$, given in Eq.~(\ref{Diffx}), agrees very well with the numerical data for all temperatures, as it is shown in Fig.~\ref{fig:D}.

\section{Expansion of the Fokker-Planck equation}
\label{sec:FP}
The non-linearity of the potential width determined by $k(x)$ thwarts any attempt for the exact solution of the  two-dimensional Fokker-Planck equation corresponding to the Langevin equations~(\ref{Langevin}). Nevertheless,  for a narrow channel, i.e., when $\varepsilon \ll 1$, one can develop a perturbative expansion for this solution.  
To this end, it is convenient to rescale the transversal coordinate as
\begin{equation} 
\eta = \frac{y}{\varepsilon}.
\end{equation} 
Then, the Fokker-Planck equation for the probability density function $p(x,\eta,t)$ of the particle position is given by 
\begin{equation} 
\label{FokkerPlanck}
\varepsilon^{2} \left( \frac{\partial p}{ \partial t}+ \frac{\partial {j}_{x}}{\partial x} \right)   
+\frac{\partial {j}_{\eta}}{\partial \eta} =0 ,  
\end{equation} 
where the two components of the probability current read 
\begin{eqnarray}
\label{CurrentComponentx}
 {j}_{x}(x,\eta,t ) =
-\left[ T_x \frac{\partial }{\partial x}  +   \frac{ k'(x) }{2 }\eta^{2} \right]p(x,\eta,t ) ,\\ 
\label{CurrentComponenteta}
 {j}_{\eta}(x,\eta,t )  = - \left[
T_y \frac{\partial }{\partial \eta}  +  k(x)\eta \right] 
p(x,\eta,t). 
\end{eqnarray} 

Let us now assume that the exact solution of the Fokker-Planck equation~(\ref{FokkerPlanck}) can be written as a series in $\varepsilon^{2}$, 
\begin{equation}
\label{expansion}
p  = p^{(0)} + \varepsilon^{2} p^{(1)} + O(\varepsilon^{4}), \qquad  
{\vec j} = {\vec j}^{(0)} + \varepsilon^{2} {\vec j}^{(1)} + O(\varepsilon^{4}). 
\end{equation} 
Individual components of the probability current $\vec{j}^{({n})}$ are defined as in Eqs.~(\ref{CurrentComponentx}) and (\ref{CurrentComponenteta}), but with the PDF $p$ being replaced by the corresponding $p^{(n)}$. Inserting expansions~(\ref{expansion}) into the Fokker-Planck equation~(\ref{FokkerPlanck}) yields
\begin{equation} 
\label{StacOr0} 
\frac{\partial j^{(0)}_{\eta}}{\partial \eta} =0, \qquad   
\left( 
 \frac{\partial p^{(n)}}{ \partial t}+ \frac{\partial j^{(n)}_{x}}{\partial x} \right)
  + \frac{\partial j^{(n+1)}_{\eta}}{\partial \eta} =0,
 \qquad n=0,1,2,\ldots 
\end{equation}
The equations (\ref{StacOr0}) must be supplemented with proper normalization and boundary conditions consistent with those for  Eq.~(\ref{FokkerPlanck}). The PDF $p$ is normalized to unity, we preserve this normalization by normalizing individual functions $p^{(n)}$ in~(\ref{expansion})  as follows,  
\begin{eqnarray}
\label{normalization}
\int_{-\infty}^{+\infty} dx \int_{-\infty}^{+\infty}d\eta\, p^{(0)}(x,\eta,t) = 1,\\ 
\int_{-\infty}^{+\infty} dx \int_{-\infty}^{+\infty}d\eta\, p^{(n)}(x,\eta,t) = 0, \qquad n=1,2,\ldots
\label{normalizationn}
\end{eqnarray}
Due to the confinement, the probability current vanishes for large $|\eta|$. This implies the boundary conditions imposed on  all terms in the expansion~(\ref{expansion}). 
In particular we assume that  
\begin{equation}
\label{CurrenBCj}
\lim_{\eta \to \pm \infty} j^{(n)}_{\eta}(x,\eta,t) =0, 
\end{equation}
holds for any $n$, $x$, and $t$. We frequently use Eq.~(\ref{CurrenBCj}) in the following derivations.

Finally, let us discuss two limitations of validity of the above perturbation expansion. Firstly, for the expansion~(\ref{StacOr0}) to be valid globally in the channel, the inequality 
\begin{equation} 
\label{smallwidth}
 {\frac{ \varepsilon^{2}}{k(x)}} \ll 1,
\end{equation}
must be satisfied {\em for any} $x$. Otherwise, the expansion would be valid only locally and it can fail badly  in extremely wide regions, where $k(x)$ is of the order of $\varepsilon$ or $\varepsilon^{2}$. For all graphical illustrations we have chosen $k(x)$ given in Eq.~(\ref{SpringConst}), and $\varepsilon^{2}$ such that the inequality (\ref{smallwidth}) holds globally in the channel. 
Secondly, from Eq.~(\ref{FokkerPlanck}) it is clear that the present perturbation expansion may fail for a large $T_x$. 
In this case, the product $\varepsilon^{2}T_x$ in Eq.~(\ref{FokkerPlanck}), arising from $\varepsilon^{2}j_x$, is no-longer small.  We will return to this point in Sec.~\ref{sec:v} when discussing the behavior of $v$ for large $T_x$ depicted in Fig.~\ref{fig:vTx}.

\section{The mean velocity} 
\label{sec:v}
\subsection{Reduced PDF and current}

The principal quantity of interest is the long-time mean velocity of the particle  $v(T_x,T_y)$ as defined in (\ref{vD_deff}). It is obtained by the integration of the {\em reduced} stationary probability current (defined below) over the unit cell of the potential \cite{Reimann},
\begin{equation}
\label{velocitydef} 
 v(T_x,T_y)  = 
\int_{0}^{1}d x  \int_{-\infty}^{+\infty}d \eta\,  J_{x}(x,\eta).
\end{equation} 

Similarly as $\vec{j}$, the reduced stationary probability current $\vec{J}=(J_x,J_\eta)$ is defined by Eq.~(\ref{CurrentComponentx}) in which, however,  the PDF  $p(x,\eta,t)$ should be replaced by the {\em reduced} stationary PDF $P(x,\eta)$. The latter solves the stationary  Fokker-Planck equation~(\ref{FokkerPlanck}), i.e., Eq.~(\ref{FokkerPlanck}) with $\partial P/\partial t =0$, subject to the conditions  
\begin{equation} 
\label{periodicityANDnormalization}
P(x+1,\eta)=P(x,\eta),\qquad
\int_{0}^{1}d x  \int_{-\infty}^{+\infty}d \eta\,  P(x,\eta) = 1. 
\end{equation} 
That is, in contrast to $p$, the reduced PDF $P$ is  periodic with the period of the potential and it is normalized to unity in the potential unit cell.
 The relations between $p$, $\vec{j}$ and the reduced quantities $P$, $\vec{J}$ are given by the summations 
\begin{equation}
\label{reducedSUM}
P(x,\eta,t) = \sum_{n=-\infty}^{+\infty } p(x+n,\eta,t),
\qquad 
\vec{J}(x,\eta,t) = \sum_{n=-\infty}^{+\infty } \vec{j}(x+n,\eta,t).
\end{equation} 
We refer to the review article \cite{Reimann} for further details.  
 
In the following we compute the first two terms, $v_0$ and $v_1$, of the expansion of the mean velocity
\begin{equation}
\label{velocityseries} 
v(T_x, T_y) =v_{0}(T_{x},T_{y} )+ \varepsilon^{2} v_{1}(T_{x},T_{y} ) + O(\varepsilon^{4}).
\end{equation}
The results are given in Eqs.~(\ref{v0})~and~(\ref{v1TxTy}). 
The individual quantities are obtained by integration of the corresponding terms of expansion of the reduced probability current,  
${\vec J} = {\vec J}^{(0)} + \varepsilon^{2} {\vec J}^{(1)} + O(\varepsilon^{4})$. 
We have  
\begin{equation} 
 v_{n}(T_x, T_y ) = \int_{0}^{1}d x  \int_{-\infty}^{+\infty}d \eta\,  J_{x}^{(n)}(x,\eta),\quad n=0,1,2,\ldots  
\end{equation} 
Hence to compute the first two terms in Eq.~(\ref{velocityseries}), we should first derive the first two terms $P^{(0)}$ and $P^{(1)}$ (or $J_x^{(0)}$ and $J_x^{(1)}$) of the expansion of the stationary reduced PDF, $ {P} = {P}^{(0)} + \varepsilon^{2} {P}^{(1)} + O(\varepsilon^{4})$ (or the current).

\subsection{Lowest-order terms $P^{(0)}(x,\eta)$ and $v_0(T_x,T_y)$}
\label{sec:lowest-order}

The leading term  $P^{(0)}$ is obtained from the stationary versions of the first two equations of the  expansion~(\ref{StacOr0}),
\begin{equation} 
\label{reducedSTAC} 
\frac{\partial J_{\eta}^{(0)}}{\partial \eta} =0, \qquad 
\frac{\partial J^{(0)}_{x}}{\partial x}  + \frac{\partial J^{(1)}_{\eta}}{\partial \eta} =0.
\end{equation} 
The first of these stationary equations is satisfied by the function $P^{(0)}$ of the form 
\begin{equation}
\label{P0A}
P^{(0)}(x,\eta) = A(x) {\rm e}^{-\frac{k(x)}{2 T_y} \eta^{2} },
\end{equation} 
with the unknown function $A(x)$, yet to be determined from the second equation in~(\ref{reducedSTAC}). 
If we integrate this second equation with respect to the coordinate $\eta$ and use the boundary condition (\ref{CurrenBCj}), $J^{(1)}_{\eta}(x,\pm \infty)=0$, we end up with the second-order ordinary differential equation for the function $A(x)$: 
\begin{equation} 
\frac{\partial }{\partial x}
\int_{-\infty}^{+\infty} d\eta\,
J^{(0)}_{x}(x,\eta)
 = 0.  
\end{equation} 
This equation is immediately integrated with respect to $x$:  
$ \int_{-\infty}^{+\infty} d\eta\, J^{(0)}_{x}(x,\eta) = v_{0}$, resulting in the first-order ordinary differential equation for $A(x)$, where $v_0$ plays the role of the integration constant (see also  Eq.\ (\ref{velocitydef})). 
In accordance with (\ref{periodicityANDnormalization}), the resulting function must be periodic, $A(x)=A(x+m)$, for any integer $m$. 
This condition implies, after some algebra, that $v_0$ is identically equal to zero,
\begin{equation} 
\label{v0}
v_{0}(T_{x},T_{y} )=0. 
\end{equation} 
Hence the ratchet effect due to $T_x \neq T_y$ is at least of the second order in the channel width. 
Finally for $A(x)$ we obtain
\begin{equation}
\label{Axi}
A(x) = \mathcal{N} \left( \sqrt{\frac{2\pi T_y }{k(x )}}
 \right)^{\frac{T_y-T_x}{T_x}},
\end{equation}  
where, the remaining normalization constant $\mathcal{N}$ is such that the PDF $P^{(0)}$ is normalized to one in the unit cell of the potential, cf.~Eq.~(\ref{periodicityANDnormalization}). 

\begin{figure}[t!]
\begin{center}
\includegraphics[width=0.95\linewidth]{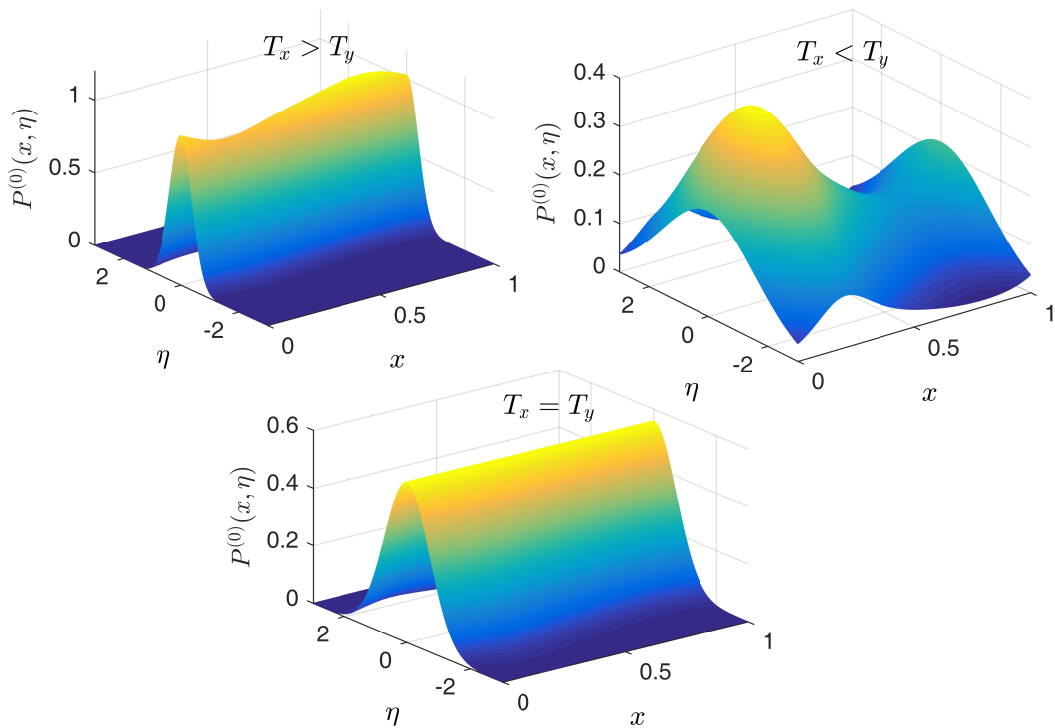}
\caption{ The lowest-order approximation $P^{(0)}$, given in Eq.~(\ref{P0A}), of the reduced stationary PDF $P$ plotted for three temperatures. The left panel: $T_x = 1$, $T_y=0.5$, the PDF is concentrated near the channel central line, it exhibits a maximum in the narrowest region of the channel. The right panel: $T_x=1$, $T_y=3$, the PDF possesses a  maximum in the widest region of the channel. The lower panel: $T_x=T_y=1$, in this case, $P^{(0)}$ is nothing but the canonical equilibrium distribution in the unit cell.}
\label{fig:P0}
\end{center}
\end{figure}

Notice that for $T_x = T_y$, the reduced PDF $P$ (and the derived function $P^{(0)}$), equals to the canonical equilibrium distribution. This is no longer true when $T_x \neq T_y$. However, the form of the PDF~(\ref{P0A}) demonstrates that, in the narrow channel ($\varepsilon \ll 1$), the Gibbs equilibrium holds locally in the transversal direction. The expression $\sqrt{\left. T_y \right/k(x)}$, which appears in both the exponent and the prefactor $A(x)$ in Eq.~(\ref{P0A}), corresponds to the local width of the distribution $P^{(0)}$. For $T_x>T_y$ the PDF $P^{(0)}$ is concentrated near the central line $\eta=0$ of the channel. On the other hand, for $T_x<T_y$ the PDF $P^{(0)}$ spreads significantly in the transversal direction. 

Physical reasons for $v_0=0$ stem from the local equilibrium form of the leading approximation $P^{(0)}$. In this order of perturbation theory, for each fixed $x$, the stationary distribution is thermalized   with the transversal heat bath at the temperature $T_y$. 
The local detailed balance then implies that there is no net heat current into the transversal heat bath, therefore also the heat current through the whole system vanishes. 
The absence of the net heat current immediately implies that the system can not act as a heat engine (or transducer), where the heat flow causes the average particle drift, and thus $v_0=0$. 

 Another intriguing and perhaps unexpected property of the PDF $P^{(0)}$ can be clearly seen in Fig.~\ref{fig:P0}. It is related to the position of the maximum of the PDF when $T_x >T_y$. The maximum determines the region, where the particle appears most often. In the well known case of thermodynamic equilibrium ($T_x = T_y $), the most probable particle position is near the minimum of the parabolic potential ($\eta=0$). In this case, for $\eta=0$ the PDF is flat in $x$-direction. 
When $T_x<T_y$, the PDF exhibits a maximum in the widest part of the channel near $x=0.25$ (cf.~equipotentials in Fig.~\ref{fig:model}).  On the other hand, when $T_x > T_y$ the maximum of the PDF is located in the \emph{narrowest region} of the channel around $x=0.75$.

In Sec.~\ref{sec:DiffusionCoeff}, when discussing the effective diffusion coefficient, we will present another argument why, to the lowest order in $\varepsilon$, the mean velocity vanishes ($v_0=0$). Let us now turn to the computation of $P^{(1)}$ and $v_1$. 

\subsection{Second terms $P^{(1)}(x,\eta)$ and $v_1(T_x,T_y)$ of the asymptotic expansions}

At first glance, the partial differential equation for $P^{(1)}$,
\begin{equation} 
\fl
 \left( 
T_x \frac{\partial^{2}}{\partial x^{2}} + \frac{\partial}{\partial x}  \frac{k'(x)}{2}\eta^{2} 
\right)P^{(0)}(x,\eta) +
\left(
T_y \frac{\partial^{2}}{\partial \eta^{2} } + \frac{\partial}{\partial \eta}k(x)\eta
\right) P^{(1)}(x,\eta)
 = 0,
 \label{stacO2} 
\end{equation} 
 is rather hard to solve (as noted e.g.~in Ref.~\cite{MartensChaos2011}, where the hard-wall corrugated channel was discussed). Fortunately, one can exploit the symmetry of the parabolic potential to get the exact expression for $P^{(1)}$. Clearly, the stationary PDF should be an even function of $\eta$. Thus one can start with the power series representation of  $P^{(1)}$ written as  
\begin{equation} 
\label{seriesP1}
P^{(1)}(x,\eta ) = \sum_{n=0}^{\infty}C_{n}(x) \eta^{2 n} 
{\rm e}^{-\frac{k(x)}{2 T_y} \eta^{2} }, 
\end{equation} 
which is not an approximation since any even function of $\eta$ can be formally written in this way.  
Now,  we insert this series and the already derived expression for $P^{(0)}$  into Eq.~(\ref{stacO2}) and collect the terms which are multiplied by the same power of $\eta$. If we require that each such term vanishes, we obtain coupled differential equations for the coefficients $C_{n}(x)$. 
After somewhat tedious algebra, it turns out that {\em only the first three coefficients} of the series (\ref{seriesP1}) are nonzero.  Moreover, only two of these coefficients, $C_{1}(x)$ and $C_{2}(x)$, follow by this approach directly from Eq.~(\ref{stacO2}):
\begin{eqnarray} 
\label{c1}
C_{1}(x) &=& - \frac{T_x}{2 T_y} \frac{{\rm d}^{2} A}{{\rm d}x^{2} },\\
\label{c2}
C_{2}(x) &=&  \frac{T_x}{8T_y^{2}}\frac{{\rm d}A}{{\rm d}x  }
\frac{{\rm d} k}{{\rm d}x },\\
C_{n}(x) &=& 0, \quad {\rm when}\quad n\geq 3 .
\end{eqnarray}

The remaining coefficient $C_{0}(x)$ is derived along the same lines as $A(x)$, that is, from the stationary equation $\partial J^{(1)}_{x}/{\partial x}  + {\partial J^{(2)}_{\eta}}/{\partial \eta} =0$, integrated over $\eta$. Imposing the boundary condition (\ref{CurrenBCj}) that the transverse current $J_{\eta}^{(2)}$ vanishes for $\eta \to \pm \infty$, we obtain
\begin{equation}  
\frac{\partial }{\partial x}
\int_{-\infty}^{+\infty} d\eta\,
J^{(1)}_{x}(x,\eta)
 = 0. 
\end{equation} 
 The integration with respect to $x$ yields the first-order differential equation for $C_{0}(x)$ which has the form 
 $\int_{-\infty}^{+\infty} d\eta\, J^{(1)}_{x}(x,\eta) = v_{1}$, where $v_1$ plays the role of the integration constant. 
 Let us now write this equation explicitly, since it is crucial for derivation of both the function $P^{(1)}(x,\eta)$ and the mean velocity $v_{1}(T_x,T_y)$. The equation reads 
\begin{equation}
\label{EQforC0}
\frac{{\rm d} }{{\rm d} x}
\left\{ C_{0}(x) \left[ M_{0}(x)\right]^{\frac{T_x -T_{y}}{T_{x}}} \right\} =
\frac{1}{T_x } 
\frac{R(x)-v_{1}}{ \left[ M_{0}(x) \right]^{T_{y}/T_x}},
\end{equation} 
where we have introduced the auxiliary known periodic functions 
\begin{equation}
\fl 
R(x) =
T_y \left[C_{1}(x) \frac{{\rm d} M_{1}}{{\rm d}x} + 
C_{2}(x)\frac{{\rm d} M_{2}}{{\rm d}x} \right] 
 - T_x \frac{\rm d}{{\rm d}x} \left[C_{1}(x)M_{1}(x)+C_{2}(x)M_{2}(x) \right]  ,
\end{equation}
and  
\begin{equation}
\label{Mn}
M_{n}(x) = \int^{+\infty}_{-\infty}{\rm d}\eta \, \eta^{2 n}\, 
{\rm e}^{-\frac{k(x)}{2 T_y} \eta^{2} } ,\qquad n=0,1,2.
\end{equation}
The solution $C_{0}(x)$ of Eq.~(\ref{EQforC0}) must satisfy $C_{0}(x+m)=C_{0}(x)$ for any integer $m$, which will be the case provided the following condition holds  
\begin{equation}
\label{r36}
0=\int_{x}^{x+1} {\rm d}x' \, 
\frac{R(x')-v_{1}}{\left[ M_0(x') \right]^{T_y/T_x}}. 
\end{equation}
The above requirement~(\ref{r36}) holds for any $x$. Choosing $x=0$, the mean velocity is given by 
\begin{equation}
\label{v1TxTy}
v_{1}(T_{x},T_{y} ) = \left. \left( 
\int_{0}^{1} {\rm d}x' \, 
\frac{R(x')}{\left[ M_0(x') \right]^{T_y/T_x}}
\right) \right/ \left( \int_{0}^{1} {\rm d}x' \, 
\frac{1}{\left[ M_0(x') \right]^{T_y/T_x}}\right) . 
\end{equation}
Finally, the solution of the Eq.~(\ref{EQforC0}) reads 
\begin{equation}
\label{C0result}
C_{0}(x) = \left( \sqrt{\frac{2\pi T_y }{k(x )}}
 \right)^{\frac{T_y-T_x}{T_x}}\left[ \mathcal{M} + 
\frac{1}{T_x} 
\int_{0}^{x} {\rm d}x' \, 
\frac{R(x')-v_{1}}{\left[ M_0(x') \right]^{T_y/T_x}} 
\right],
\end{equation}
where the integration constant $\mathcal{M}$ is chosen in accordance with the normalization condition~(\ref{normalizationn}) such that
\begin{equation}
\int_{0}^{1}d x \int_{-\infty}^{+\infty}d \eta 
\,  P^{(1)}(x,\eta) = 0. 
\end{equation}

\begin{figure}[t!]
\begin{center}
\includegraphics[width=1.0\linewidth]{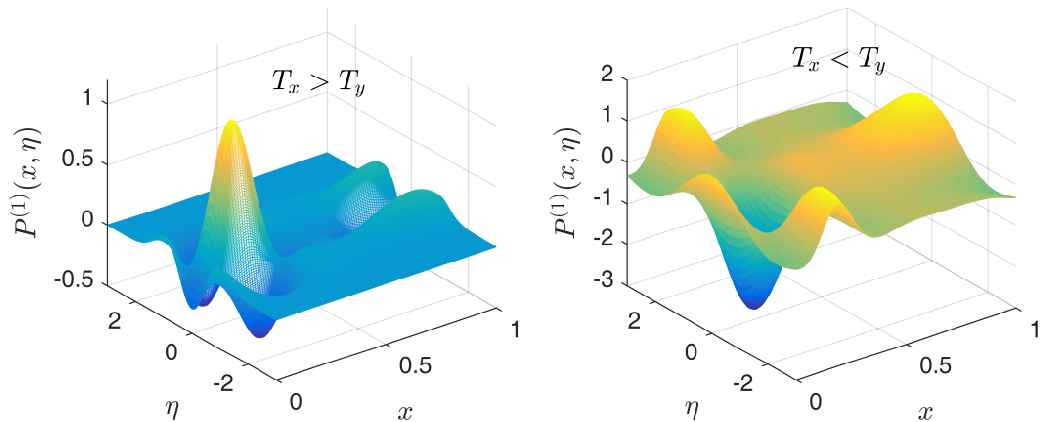}
\caption{The function $P^{(1)}$~(\ref{seriesP1}) plotted for two different temperatures $T_y$ (the left panel: $T_x = 1$, $T_y=0.5$; the right panel: $T_x=1$, $T_y=3$). The effect of the correction $\varepsilon^{2} P_1$ on the PDF $P$ is that the maximum of the principal term $P^{(0)}$, shown in Fig.~\ref{fig:P0}, is reduced and the side-peaks tend to appear.}
\label{fig:P1}
\end{center} 
\end{figure}

\begin{figure}[t!]
\begin{center}
\includegraphics[width=1\linewidth]{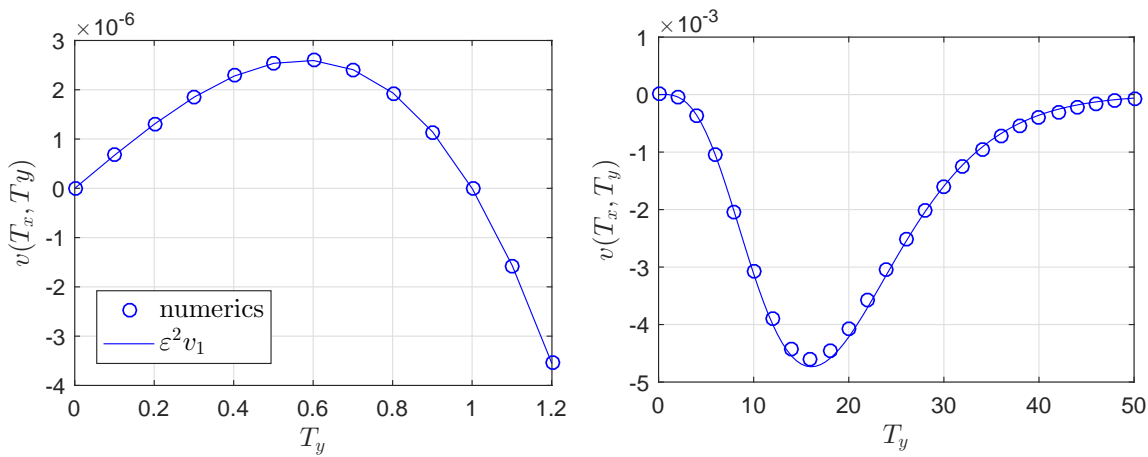}
\caption{Mean velocity of the particle as the function of $T_y$  for $T_x=1$. Analytical curves (solid lines) are plotted using Eq.~(\ref{v1TxTy}). In the numerics ({\color{blue}$\bigcirc$}) we set $\varepsilon=0.01$. The left panel magnifies the region $T_x>T_y$, which is not well visible in the right panel. }
\label{fig:vTy}
\end{center}
\end{figure} 

Derived coefficients $C_{n}(x)$, given in Eqs.~(\ref{c1}), (\ref{c2}) and (\ref{C0result}),  when inserted back into the Ansatz~(\ref{seriesP1}), yield exact expression for $P^{(1)}$. In Fig.~\ref{fig:P1} we plot this function for $T_x < T_y$ and $T_x > T_y$.\footnote{When $T_x = T_y$, the PDF $P$ is given by the Gibbs equilibrium distribution. It equals exactly to $P^{(0)}$ and all higher-order corrections $P^{(n)}$ vanish.} 
Notice that the principal part determining an overall global shape of the exact PDF $P$ is contained in $P^{(0)}$. The small correction $\varepsilon^{2} P^{(1)}$ describes local deviations of $P$ from $P^{(0)}$ when $T_x\neq T_y$. For both, $T_x < T_y$ and $T_x > T_y$, the maximum of $P^{(0)}$ is slightly reduced (negative $P^{(1)}$ near the maximum of $P^{(0)}$) and the two side-peaks tend to emerge. Despite its smallness, the correction is crucial for a proper description of the particle dynamics. The transversal current $J_{\eta}^{(1)}$, obtained after substitution of $P^{(1)}$ into the definition~(\ref{CurrentComponenteta}), does  not vanish when $T_x \neq T_y$. This is in sharp contrast to what we have observed in the lowest-order approximation, where $P^{(0)}$ satisfies the local detailed balance condition with the transversal heat bath and $J_{\eta}^{(0)}=0$ everywhere in the unit cell.

\begin{figure}[t!]
\begin{center}
\includegraphics[width=1\linewidth]{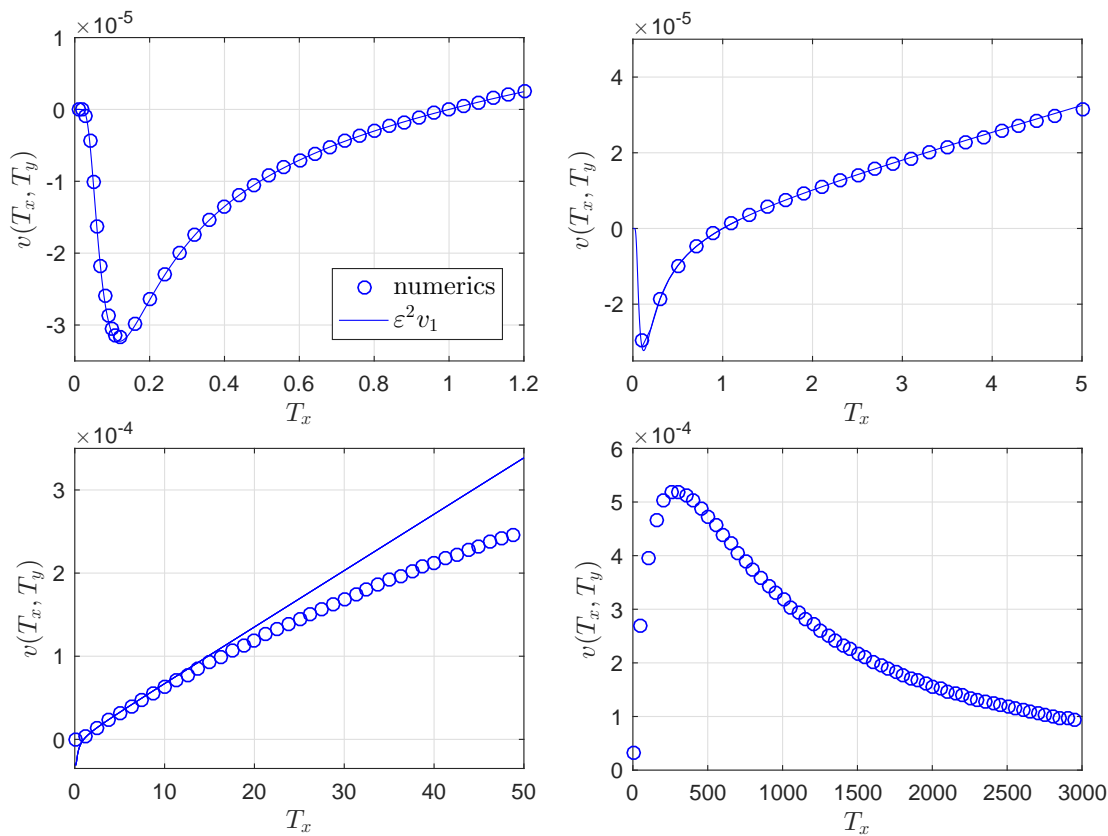}
\caption{Mean velocity of the particle as the function of $T_x$ for $T_y=1$. Analytical curves $\varepsilon^{2}v_1$ (solid lines) are plotted using Eq.~(\ref{v1TxTy}). In the numerics ({\color{blue}$\bigcirc$}) we set $\varepsilon=0.01$. The four panels differ by the range of $T_x$ only. }
\label{fig:vTx}
\end{center}
\end{figure}

The deviation from the local detailed balance with the transversal heat bath, as described by the $\varepsilon^{2} P^{(1)}$ term, allows the heat current to flow through the system. Part of the heat current now can be used to systematically transport the particle with the mean drift velocity $\varepsilon^{2} v_{1}$, where $v_{1}$ is  given in Eq.~(\ref{v1TxTy}). Figs.~\ref{fig:vTy},~\ref{fig:vTx} demonstrate a rich non-linear behavior of $v_1$ as the function of temperatures $T_y$ and $T_x$, respectively. Note that the both temperatures influence the analytical result for $v_1$ in a rather different manner. 

When $v_1$ is plotted as the function of $T_y$ (Fig.~\ref{fig:vTy}),   there exists an optimal value of $T_y$ for both $T_x>T_y$ and $T_x<T_y$. The maximum for $T_x<T_y$ is by three orders of magnitude larger than that in the region $T_x>T_y$. This disparity between $T_x<T_y$ and $T_x>T_y$ cases is clearly visible on the level of individual trajectories depicted in Fig.~\ref{fig:trajectories}, where the average drift is evident only for $T_x<T_y$ case (the right panel in Fig.~\ref{fig:trajectories}). 
In the low-temperature limit $T_y \to 0$, the transversal fluctuations supply a negligible amount of energy and hence the particle diffuses close to the channel central line. In this limit we  observe a free one-dimensional Brownian motion with $v=0$. 
On the other hand, the larger $T_y$ the farther the particle goes in the transversal direction  spending considerable amount of time in the widest part of the unit cell. Increasing $T_y$ eventually reduces the mean particle velocity to zero. 
The derived analytical expression~(\ref{v1TxTy}) for $v_1$ is in a reasonable agreement with the numerical data {\em in the whole temperature range} $T_y \in (0,\infty)$. We can conclude that the perturbation expansion captures all essential physical behavior for any $T_y$ and a fixed $T_x$ (provided that $\varepsilon^{2}/k(x) \ll 1$ holds for any $x$, and $\varepsilon^{2}T_x$ is small, cf.~Eq.~(\ref{smallwidth})).

The situation is rather different when we plot the result (\ref{v1TxTy}) as the  function of $T_x$  (Fig.~\ref{fig:vTx}). The analytical expression for $v_1$ exhibits an extreme when $T_x<T_y$, whereas for $T_x>T_y$ it grows linearly for large $T_x$. The latter behavior is physically incorrect and hence the analytical result for $v_1$ differs from the numerical data. 
Numerical analysis predicts correctly that $v_1$ should slowly approach zero as $T_x \to \infty$. Why does the expansion differs so drastically from reality in this case? The lowest-order term $P^{(0)}$ as given in~Eq.~(\ref{P0A}) embodies the assumption that {\em the local equilibrium with the transversal heat bath at the temperature $T_y$ holds} in the lowest order of the theory. However, this is no longer true in the limit $T_x\to \infty$, where the lowest-order term should actually represent the fact that the local equilibrium holds with the {\em longitudinal} heat bath at the temperature $T_x$. 
Hence, to describe the limit $T_x\to \infty$ properly, one should devise a different perturbation expansion with the small parameter $1/T_x$, $1 / T_x \ll 1$.\footnote{In fact, to the lowest order in $1/T_x$, the reduced stationary PDF equals to $ \exp\!\left( - \frac{\overline{k}}{T_y} y^{2} \right)/Z$, where $\overline{k}=\int_{0}^{1}dx\,k(x)$, and $Z$ is a normalization. The asymmetry of the channel walls is thus averaged out and no ratchet effect occurs in the limit $T_x \to \infty$.}
In  Sec.~\ref{sec:DiffusionCoeff} we will offer another physically motivated argument for the fact that indeed $v\to 0$ as $T_x \to \infty$.

\section{The effective diffusion coefficient}
\label{sec:DiffusionCoeff}

The principal part $D_0$ of the effective longitudinal diffusion coefficient $D$, defined in~Eq.~(\ref{vD_deff}) and expanded according to Eq.~(\ref{Dexpanded}), can be derived directly from the lowest-order Fick-Jacobs approximation \cite{Zwanzig1992, RegueraRubi, CPHC:CPHC200800526}. The $\varepsilon$-expansion~(\ref{StacOr0}) of the two-dimensional Fokker-Planck equation~(\ref{FokkerPlanck}) separates time-scales of the transversal and the longitudinal dynamics. In the lowest order in $\varepsilon$ this separation yields a closed  diffusion equation for the marginal longitudinal PDF $p^{(0)}(x,t)$, $p^{(0)}(x,t)=\int_{-\infty}^{+\infty}d\eta \, p^{(0)}(x,\eta,t)$. The equation is known as the Fick-Jacobs equation \cite{Zwanzig1992, RegueraRubi, CPHC:CPHC200800526} and it is obtained by $\eta$-integration of the $n=0$ term of the expansion~(\ref{StacOr0}). The result reads  
\begin{equation} 
\label{FickJacobs}
\frac{\partial} {\partial t}p^{(0)}(x,t) = T_x \frac{\partial^{2}}{\partial x^{2}} p^{(0)}(x,t) + 
\frac{\partial}{\partial x} \left[ \mathcal{F}'(x) p^{(0)}(x,t) \right], 
\end{equation}
where the effective potential $\mathcal{F}(x)$ is given by the expression resembling the equilibrium free energy: 
\begin{equation}
\mathcal{F}(x)= -T_y \log\left[ M_{0}(x) \right].
\end{equation}

\begin{figure}[t!]
\begin{center}
\includegraphics[width=0.9\linewidth]{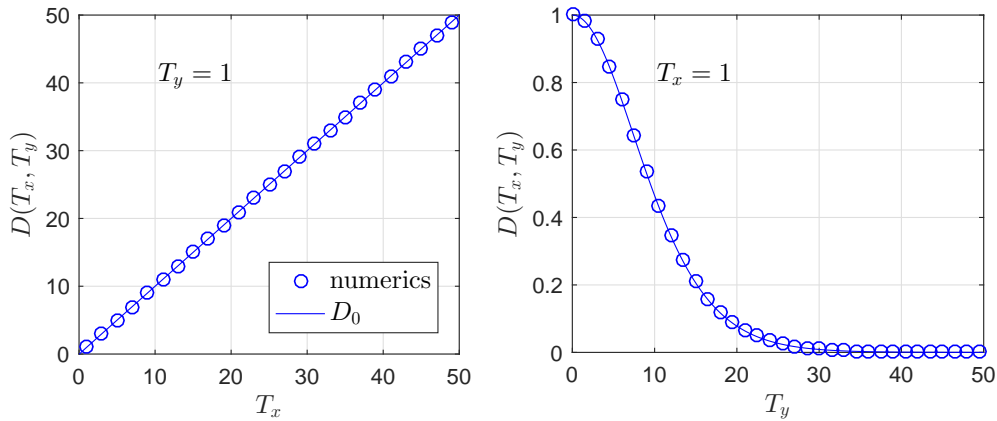}
\caption{Lowest-order approximation $D_0$~(\ref{Diffx}) (solid lines) of the effective longitudinal diffusion coefficient $D$, compared to the numerics ({\color{blue}$\bigcirc$}) for different temperatures.}
\label{fig:D} 
\end{center}
\end{figure}

Since the ``local partition function'' $M_{0}(x)$, defined in Eq.~(\ref{Mn}), is a periodic function of $x$, the Fick-Jacobs equation~(\ref{FickJacobs}) describes the diffusion in the  periodic potential $\mathcal{F}(x)$. For such a diffusion process, the asymptotic mean particle velocity vanishes, which is consistent with the result  $v_0=0$ obtained in the preceding section, cf.~Eq.~(\ref{v0}). 

The effective diffusion coefficient for the one-dimensional diffusion in the periodic potential $\mathcal{F}(x)$ is given by \cite{LifsonJackson, FESTA1978229, ReimannBroeckI,ReimannBroeckII}
\begin{equation}
\label{Diffx}
D_0(T_x,T_y) = 
\frac{T_x}{\left( \int_{0}^{1}dx \left[ {k(x)} \right]^{\frac{T_y}{2T_x}}  \right)
\left( \int_{0}^{1} dx \left[ {k(x)} \right]^{-\frac{T_y}{2T_x}}  \right)}.
\end{equation}
See also Ref.~\cite{Vulpiani2014}, where various known corrections to this result are summarized and discussed. In our case it is sufficient to use the classical form~(\ref{Diffx}) which follows directly from the formula obtained by Lifson and Jackson \cite{LifsonJackson}. 
Eq.~(\ref{Diffx}) captures the leading behavior of the effective diffusion coefficient. 
The first-order correction $\varepsilon^{2}D_1$ to this behavior  is vanishingly small as compared to $D_0$. 
In Fig.~\ref{fig:D}, we compare the leading term $D_0$ given by Eq.~(\ref{Diffx}) with the numerically obtained values of $D$ for different temperatures $T_x$, $T_y$. 

 When the temperature $T_x$ is high enough, the concrete form of the confining potential plays a minor role and the particle effectively diffuses along a {\em straight} tube.  Thus the effective diffusion coefficient approaches that of the free particle diffusing in one dimension, $D_{free}=T_x$ (the left panel in Fig.~\ref{fig:D}) and the mean particle velocity vanishes (Fig.~\ref{fig:vTx}) since the particle feels no asymmetry of the potential. Returning to the mechanical ratchet in Fig.~\ref{fig:model}, in the limit $T_x\to \infty$, the wheels fluctuate so violently that their effect averages out and the pawl does not feel the asymmetry of the teeth.  
 The inhomogeneities of the confining potential manifest themselves for small values of $T_x$, where the values of $D$ are below $D_{free}$. 

For a fixed $T_x$, the effective diffusion coefficient is significantly reduced with increasing $T_y$ (the right panel in Fig.~\ref{fig:D}). For large $T_y$ the particle spends substantial amount of time wandering in the widest part of the potential. Only occasionally it enters the narrow region and passes through it to a neighboring potential unit cell. This dynamical trapping becomes more pronounced for larger $T_y$. The effective diffusion coefficient $D$ eventually approaches zero as $T_y\to \infty$. In the mechanical ratchet (Fig.~\ref{fig:model}), this limit means rapid fluctuations of the pawl, which effectively stalls the slowly rotating wheels.

\section{Concluding remarks and perspectives}
\label{sec:conclusion}

In the present work, we have developed the systematic expansion of the Fokker-Planck equation which goes beyond the standard Fick-Jacobs approximation. The expansion has been used to analyze dynamics and performance of the two-dimensional stochastic model of the Feynman ratchet sketched in Fig.~\ref{fig:model}. In the model, the ratchet effect is induced by the simultaneous coupling of the Brownian particle to two heat baths at different temperatures. Notably, the ratchet effect cannot be captured by the standard Fick-Jacobs approximation (the lowest order of our expansion), where the system is equilibrated with the transversal heat bath and consequently no heat flows between the two reservoirs (as discussed at the end of Sec.~\ref{sec:lowest-order}). The symmetry of the confining potential~(\ref{potentialGEN}) allowed us to derive also the second-order term of the expansion which provided satisfactory explanation of  the working principle of the ratchet.  

In particular, our approach yields analytical expressions for both the mean velocity $v$ of the rotation of ratchet wheels  (depicted in Figs.~\ref{fig:vTy} and~\ref{fig:vTx}) and  the effective diffusion coefficient $D$ (see Fig.~\ref{fig:D}) which describes how noisy this rotation is. It also provides a refined information about the position of the pawl within the potential unit cell (reduced stationary PDFs plotted in Figs.~\ref{fig:P0}). For $T_x=T_y$ the PDF equals to the canonical equilibrium distribution. For $T_x\neq T_y$ it exhibits peaks in the widest part (when $T_x < T_y$) or, surprisingly, in the narrowest part (when $T_x > T_y$) of the unit cell. The fine structure of these peaks, which actually gives rise to the ratchet effect, is revealed by the first-order correction $P^{(1)}$ plotted in Fig.~\ref{fig:P1}.

From a general perspective, the present paper demonstrates that expansions inspired by  the Fick-Jacobs theory can be used successfully to study transport phenomena in inherently two-dimensional ratchets coupled simultaneously to several heat reservoirs. Leaving aside straightforward generalizations such as addition of more dimensions, heat baths, and external drifts (or loads), there are several intriguing non-trivial directions for the extensions of the present theory. For instance, it is definitely a challenging task to derive transport coefficients for many-dimensional {\em inertial} ratchets diffusing in the environment with inhomogeneous temperature distribution (i.e., for many-dimensional generalizations of the B{\" u}ttiker-Landauer ratchets \cite{Seifert2012, Sekimoto1997, Buttiker1987, Landauer1988, AstumianPRL1999, AstumianPRE1999, HondouSekimotoPRE2000, Ai2005, Asfaw2005}). For such models, even in one dimension it is known that the overdamped limit is singular, i.e., neglect of the inertial effects significantly changes working principle of the ratchet \cite{Sekimoto1997}. Study of these ratchets in several dimensions could uncover new effects arising from coupling of different non-equilibrium processes. 

Another direction of research, which stems from the need to understand actual intracellular transport, concerns  ratchets with many interacting particles  \cite{SlaninaEPL2008, SlaninaJSP2009, SlaninaPRE2009}.  A generally accepted extension of the Fick-Jacobs theory for interacting particles does not exist yet (see Refs.~\cite{Bruna2012, Bruna2014, SuarezPRE2015, SuarezMartinPRE2015, Martin2016} for recent studies). A systematic expansion in the channel width, performed on simple models similar to that studied in the present work, could shed some light on this interesting topic. One may expect that the lowest-order approximation would be similar to the single-file dynamics \cite{Chaudhury2015PRE, RC2014PRE, RC2012JCP, Diego2014} and that the higher-order corrections would describe a subtle interplay between interaction and the channel asymmetry, which can affect performance of the ratchet in an unexpected way.

\ack
Authors are grateful to Pavol Kalinay for illuminating discussions regarding generalizations of the Fick-Jacobs theory and to the organizers of XXV Sitges Conference on Statistical Mechanics for creating a stimulating environment for such debates.  
MHY is thankful to the Charles University in Prague for their hospitality and financial support during his visit in summer 2015. MHY also acknowledges the guidance and help of the International Cooperation Scientific Department of Zanjan University.

\appendix
\section{Numerical calculation of transport coefficients in two dimensions}
\label{App:numerics}

\subsection{Calculation of the stationary reduced PDF and the mean velocity}

To compare the results of the perturbative theory with the exact solution, we discretize the Fokker-Planck equation (\ref{FokkerPlanck}) in a unit cell of the potential and solve numerically the resulting stationary Master equation. The result of the numerical procedure is the probability ${\mathcal P}_{ij}$ to find the particle near the grid point $[i,j]$, $i=i(x)$, $j=j(y)$, which approximates the stationary reduced PDF $P(x, y)$, for definitions see Eqs.~(\ref{eq:ixjy})-(\ref{eq:reduced_PDF_num}).

The numerical solution can be found in a finite region only. In the $x$-direction this region is naturally determined by one period of the potential, we take $x \in [0,1)$. In the $y$-direction there is no such natural limitation. Nevertheless, for confining potentials like the potential in Eq.~(\ref{potentialGEN}), it is plausible to introduce a cutoff $Y$ such that the probability to find the particle at the boundary $|y|=Y$ near the cutoff is small, and hence one can set ${\mathcal P}_{ij(y)} = 0$ for $|y| > Y$. In the implementation of the numerical procedure, the $Y$ value can be determined iteratively (the Master equation can be solved for arbitrary $Y$) such that the probability to find the particle near the cut off (for an arbitrary $x$) is smaller than a given number $p_Y$, i.e. $\max_{i} {\mathcal P}_{ij(|Y|)} < p_Y$. A reasonable choice of this number, which we use in our implementation, is $p_Y = p_{\rm max}/10^6$, where $p_{\rm max} = \max_{i,j} {\mathcal P}_{ij}$.

The condition ${\mathcal P}_{ij(y)} = 0$ for $|y| > Y$ is most easily fulfilled by a  redefinition of the potential as
\begin{equation}
V_{i(x),j(y)} = U(x,y)\,\theta(Y-|y|) + \infty\, \theta(|y|-Y)\,,
\label{eq:potential_numerics}
\end{equation}
where $\theta(x)$ stands for the unit step function. Let us discretize the region $[0,1)\times [-Y,Y]$ on $N_x+1$ sections in the $x$-direction and $N_y+1$ sections in the $y$-direction. Then we can write
\begin{equation}
i(x) = \left\lfloor\frac{x}{\Delta_x}\right\rfloor\,,
\qquad
j(y) = \left\lfloor\frac{y + Y}{\Delta_y}\right\rfloor\,,
\label{eq:ixjy}
\end{equation}
where $\Delta_x = 1/(N_x+1)$, $\Delta_y = 2Y/N_y$ and the expression $\left\lfloor f\right\rfloor$ mean that $f$ is rounded towards negative infinity. From Eqs.~(\ref{eq:ixjy}) and the conditions $x\in[0,1)$ and ${\mathcal P}_{ij(y)} = 0$ for $|y| > Y$, we inspect that the probabilities ${\mathcal P}_{i(x)j(y)}$ can assume nonzero values for $i = 0,\dots,N_x$ and $j=0,\dots,N_y$. With these definitions the stationary reduced PDF is given by
\begin{equation}
P(x,y) = \lim_{\Delta_x,\Delta_y \to 0} \frac{{\mathcal P}_{i(x)j(y)}}{\Delta_x\Delta_y}\,.
\label{eq:reduced_PDF_num}
\end{equation}
 For finite $\Delta_x$ and $\Delta_y$ the stationary Fokker-Planck equation (\ref{FokkerPlanck}) for $P(x,y)$ can be discretized as
\begin{eqnarray}
0 &=& \frac{d}{dt}{\mathcal P}_{ij} =
W_{i+1i,j} {\mathcal P}_{i+1j} +
W_{i-1i,j} {\mathcal P}_{i-1j} +
W_{i,j+1j} {\mathcal P}_{ij+1} +
\nonumber\\&+&
W_{i,j-1j} {\mathcal P}_{ij-1} -
(W_{ii+1,j} + W_{ii-1,j} + W_{i,jj+1} + W_{i,jj-1}){\mathcal P}_{ij}
\,,
\label{eq:num_masterEQ1}
\end{eqnarray} 
where the transition rates are given by
\begin{eqnarray}
W_{i_a i_b,j} = \frac{T_x }{\Delta_x^{2}} \exp\! \left[ - \frac{V_{i_bj}-V_{i_a j}}{2 T_{x}} \right]\,,
\label{eq:rates_x}
\\
W_{i,j_aj_b} = \frac{T_y }{\Delta_y^{2}} \exp\! \left[ - \frac{V_{ij_b}-V_{ij_a}}{2 T_{y}} \right]\,, 
\label{eq:rates_y} 
\end{eqnarray}
and where we impose the periodic boundary conditions in the $x$-direction: $i = 0 = N_x + 2$, $i = -1 = N_x + 1$. The advantage of this discretization is that the stationary Master equation (\ref{eq:num_masterEQ1}) yields physically reasonable steady-state distribution even for large discretization steps $\Delta_x$ and $\Delta_y$ (as compared e.g., to the finite element method, which usually generates unphysical results for large $\Delta_x$, $\Delta_y$). In the present model, transitions in the longitudinal (transversal) direction are activated by the heat reservoir with the temperature $T_x$ ($T_y$). This is ensured  by the fact that the transition rates (\ref{eq:rates_x}) and (\ref{eq:rates_y}) fulfill local detailed balance conditions with respect to corresponding Gibbs equilibrium distributions: 
$W_{i_a i_b,j} \exp(-{V_{i_aj}}/{T_x} )= W_{i_b i_a,j} \exp(-{V_{i_bj}}/{T_x} )$, 
and 
$W_{i,j_a j_b} \exp(-{V_{ij_a}}/{T_y} )= W_{i, j_b j_a} \exp(-{V_{ij_b}}/{T_y} )$. 
The system of equations (\ref{eq:num_masterEQ1}) transforms into Eq.~(\ref{FokkerPlanck}) for the reduced PDF $P(x,y)$ in the limit $\Delta_x,\Delta_y \to 0$ (or, equivalently, $N_x,N_y \to \infty$). For the potential (\ref{potentialGEN})-(\ref{SpringConst}) used for the illustrations throughout this paper, a good approximation of $P(x,y)$ is obtained already for $N_x=N_y=100$.

In order to solve the system of equations (\ref{eq:num_masterEQ1}) numerically it is favorable to rewrite it in a matrix form 
\begin{equation}
\hat{L}\mathbf{p} = \frac{d}{dt}\mathbf{p} = 0\,.
\label{eq:Lp}
\end{equation}
While this can be done in many equivalent ways, we suggest the mapping where the elements $p_k$, $k=1,\dots,(N_x+1)(N_y+1)$, of the vector $\mathbf{p}$ read 
\begin{equation}
p_{j(N_x+1) + i + 1} = {\mathcal P}_{ij}\,.
\label{eq:pi}
\end{equation}
With this definition, the form of the elements of the matrix $\hat{L}$ can be deduced from Eq.~(\ref{eq:num_masterEQ1}) in a straightforward way. The result is rather long and we will not write it here. The resulting matrix $\hat{L}$ has $(N_x+1)^2(N_y+1)^2$ elements and thus, for small $\Delta_x$ and $\Delta_y$, it is very large. Fortunately, it is also sparse, because at maximum only 5 elements in each of its row are nonzero.

This property allows us to find the stationary distribution of the system with extreme accuracy (taking values of $N_x$ and $N_y$ as large as 1500 leading to matrices $\hat{L}$ with $1501^4\approx 5\times 10^{12}$ elements). In Matlab, we recommend to use the function \emph{eigs}($\hat{L}$,1,'lr') (the arguments 1 and 'lr' tell eigs to return eigenvector of $\hat{L}$ corresponding to its smallest eigenvalue, i.e. to find the stationary distribution) together with representation of the matrix $\hat{L}$ in a computer using the \emph{sparse} command. 
Numerical solution of Eq.~(\ref{eq:num_masterEQ1}) for large $N_x$ and $N_y$ gives very good approximation of the correct $P(x,y)$ which can be used to calculate the probability current through the system. Let us note that the procedure described in this section can be used to determine a steady state for an arbitrary two dimensional problem.
Here, we have used this procedure to calculate the data for the mean velocity depicted in Figs.~\ref{fig:vTy} and \ref{fig:vTx}.

\subsection{Calculation of the effective diffusion coefficient}

In order to calculate the effective diffusion coefficient $D$ the knowledge of the stationary reduced PDF $P(x,y)$ is not sufficient. To this end we will investigate a long-time behavior of the characteristic function for the particle displacement, which is closely related to $D$. The diffusion coefficient is defined as $D = \lim_{t\to\infty}(\left< x(t)^2\right>-\left< x(t)\right>^2)/(2t)$, cf.\ Eqs.~(\ref{vD_deff}), where the average is taken over all possible values of the particle position at time $t$, $x(t)$. Further, without loss of generality, we assume that the particle departed with unit probability at time 0 from $x(0) = 0$. Defining the displacement of the particle by 
\begin{equation} 
q(t) = \int_0^t\,dt'\dot{x}(t'),
\label{eq:int_current}
\end{equation}
the diffusion coefficient can be written as 
\begin{equation}
D = \lim_{t\to\infty}\frac{\left< q(t)^2\right>-\left< q(t)\right>^2}{2t}.
\label{eq:Diffusion_coeff}
\end{equation}
Here and below the average is taken over all possible values of the displacement $q(t)$. Let us now define the characteristic function for this quantity:
\begin{equation}
\chi(u,t) = \left< \exp(-u q(t)) \right>.
\end{equation}
For long times $t$, this function yields the stationary velocity of the particle (\ref{velocitydef}) and the diffusion coefficient (\ref{eq:Diffusion_coeff}). We have
\begin{equation} 
\lim_{t\to\infty} \frac{1}{t}\frac{\partial}{\partial u}\ln \chi(u,t) \bigg|_{u=0} = v,\qquad \lim_{t\to\infty} \frac{1}{2t}\frac{\partial^2}{\partial u^2}\ln \chi(u,t) \bigg|_{u=0} = D. 
\label{eq:diff_coef_num}
\end{equation} 
In order to evaluate the last two expressions one can draw inspiration from the large deviation theory \cite{Touchette20091}. Before the state vector $\mathbf p$ reaches the steady state $d \mathbf p/dt = 0$, its dynamics is determined by the Master equation $d \mathbf p(t)/dt = \hat{L}\mathbf p(t)$ and for an infinitesimal time evolution we get $\mathbf p(t+dt) = (1+ \hat{L}dt)\mathbf p(t)$. The matrix $\hat{R} = 1+ \hat{L}dt$ is thus an infinitesimal propagator of the state vector ${\mathbf p}(t)$ and, for a finite time $\tau$, we can formally write $\mathbf p(\tau) = \hat{R}^{\tau/dt}\mathbf p(0)$. 

In a similar manner it is possible to derive an expression for the characteristic function of the particle displacement~(\ref{eq:int_current}). The displacement of the particle during an infinitesimal time interval $dt$ is $q(dt) = dtdx/dt = x(t+dt)-x(t)$. In the approximate discrete model, whenever the particle moves in the $x$-direction it overcomes the distance $\Delta_x$. Thus if it moves to the right, we have $q(dt) = \Delta_x$. Similarly, we have $q(dt) = - \Delta_x$ if the particle moves to the left, and $q(dt) = 0$ otherwise. This means that the {\em joint} probability $[\mathbf{r}(q,dt)]_i$ that the displacement generated during the small time interval $dt$ equals to $q$ and that the particle is in the state $i$ at time $dt$ can be written as $[\mathbf r(q,dt)]_i = [\hat{G}(q) \mathbf r(0,0)]_i$, with the new propagator $\hat{G}(q) = 1 + \hat{K}(q)dt$. Here, the matrix $\hat{K}(q)$ is obtained from the rate matrix $\hat{L}$ if one multiplies all its diagonal elements by $\delta(q)$ and substitutes the transition rates
\begin{eqnarray} 
W_{i_a i_b,j} = \frac{T_x }{a^{2}} \exp\! \left[ - \frac{V_{i_bj}-V_{i_a j}}{2 T_{x}} \right] \delta[q-(i_b-i_a)\Delta_x],
\label{eq:rates_x_q}
\\
W_{i,j_aj_b} = \frac{T_y }{a^{2}} \exp\! \left[ - \frac{V_{ij_b}-V_{ij_a}}{2 T_{y}} \right] \delta(q), 
\label{eq:rates_y_q}
\end{eqnarray}
for the rates (\ref{eq:rates_x})-(\ref{eq:rates_y}) contained in all off-diagonal elements of the matrix $\hat{L}$. 

The joint probability vector ${\mathbf r}(q,2dt)$ after two small time intervals is given by the convolution $\mathbf r(q,2dt) = \int_{-\infty}^{\infty}dq'\hat{G}(q-q')\hat{G}(q')\mathbf r(0,0)$. For longer time intervals $\tau$ ($\tau$ is a large multiple of $dt$) we thus end up with a multiple convolution. Taking the two-sided Laplace transform of $\mathbf r(q,\tau)$ with respect to $q$ we obtain $\tilde{\mathbf r}(u,\tau) = [\tilde{G}(u)]^{\tau/dt}\mathbf p(0)$, where we have used the transformed $\tilde{G}(u) = 1+\tilde{K}(u)dt$ with $\tilde{K}(u) =\int_{-\infty}^{\infty}dq \exp(-qu) \hat{K}(q)$. The transformed matrix $\tilde{K}(u)$ can be obtained from the original matrix $\hat{K}(q)$ by substituting $1$ for $\delta(q)$ and $\exp[-u (i_b-i_a)\Delta_x]$ for $\delta[q-(i_b-i_a)\Delta_x]$. The transformed joint probability vector $ \tilde{\mathbf r}(u,\tau)$ yields the characteristic function $\chi(u,\tau)$ via the summation of its elements over all states, $\chi(u,\tau) = \sum_i [\tilde{\mathbf r}(u,\tau)]_i$. The last expression can be formally written as $\chi(u,\tau) = \sum_i \left[\tilde{G}(u)^{\tau/dt}\mathbf p(0)\right]_i$.  

Let us now assume that the matrix $\tilde{G}(u)$ can be diagonalized as $\tilde{G}(u) = \tilde{U}_l(u) \tilde{D}(u) \tilde{U}_r(u)$, where $\tilde{U}_r(u) \tilde{U}_l(u) = 1$ and $\tilde{D}(u)$ is a diagonal matrix with eigenvalues $\tilde{d}_i(u)$ of the matrix $\tilde{G}(u)$ on its diagonal. Since $\tilde{G}(u) = 1 + \tilde{K}(u)dt$, each of these eigenvalues can be written as $\tilde{d}_i(u) = 1+\lambda_i(u) dt$, where $\lambda_i(u)$ are eigenvalues of the matrix $\tilde{K}(u)$. 
The characteristic function $\chi(u,\tau)$ can be expressed as $\chi(u,\tau) = \sum_i \left[\tilde{U}_l(u) \tilde{D}(u)^{\tau/dt}\tilde{U}_r(u) p(0)\right]_i = \sum_i c_i(u) [1+\lambda_i(u) dt]^{\tau/dt}$, where the specific form of the coefficients $c_i(u)$ turns out to be unimportant for the further calculation. The time interval $dt$ is vanishingly small as compared to $\tau$ and thus the only significant contribution to the sum is given by the summand that contains the largest eigenvalue of the matrix $\tilde{K}(u)$. Let us denote this eigenvalue simply as $\lambda(u)$ and the corresponding coefficient as $c(u)$. Then we have $\chi(u,\tau) \approx c(u) [1+\lambda(u)dt]^{\tau/dt}$. Substituting the last expression into Eq.~(\ref{eq:diff_coef_num}) we get
\begin{equation}
v = \frac{d}{du}\lambda(u)\bigg|_{u=0}\,\qquad D = \frac{1}{2}\frac{d^2}{du^2}\lambda(u)\bigg|_{u=0}.
\label{eq:diff_coef_num2}
\end{equation}

The largest eigenvalue $\lambda(u)$  of the \emph{tilted} \cite{Touchette20091} rate matrix $\tilde{K}(u)$ can be calculated numerically. In Matlab we have again used the function \emph{eigs} ($\tilde{K}(u)$ must be defined as a sparse matrix). However,  in case of a fine discretization the numerical procedure becomes unstable.
In order to circumvent this issue we have exploited the close relationship between the displacement (\ref{eq:int_current}) and the particle current. Namely, one can calculate the two averages of the displacement in Eq.~(\ref{eq:diff_coef_num}) also using the largest eigenvalue of a simpler matrix $\tilde{K}_{ii+1}(u)$ instead of $\tilde{K}(u)$. The matrix $\tilde{K}_{ii+1}(u)$ contains smaller number of tilted elements than $\tilde{K}(u)$.  Namely the transition rates $W_{i i+1,j}$, $j = 0,\dots,N_y$, to the right are tilted by $\exp[-u(i+1-i)\Delta_x(N_x + 1)]$ = $\exp(- u)$, the transition rates $W_{i+1 i,j}$, $j = 0,\dots,N_y$, to the left are tilted by $\exp[-u(i-i-1)\Delta_x(N_x + 1)]$ = $\exp(u)$ and other transition rates are the same as in the original rate matrix $\hat{L}$. The matrix $\tilde{K}_{ii+1}(u)$ thus describes the particle current just between the states with $x$-coordinates given by $i$ and $(i+1)$, but this current is multiplied $(N_x+1)\times$ with respect to the current which follows from the original matrix $\tilde{K}(u)$. Due to the continuity of the current in steady states it does not matter whether we calculate $(N_x+1)\times$ current at $1$ fixed $x$ position (what is done using $\tilde{K}_{ii+1}(u)$) or $1\times$ current at $(N_x+1)$ $x$ positions (what is done using $\tilde{K}(u)$). Differently speaking, the original tilted matrix $\tilde{K}(u)$ describes the particle displacement counted in units of the lattice constant $\Delta_x$, while the simplified tilted matrix $\tilde{K}_{ii+1}(u)$ yields large deviation properties of the `renormalized' particle displacement with increments equal to $+1$ ($-1$) which are added when the particle surpasses the distance equal to the length of the potential unit cell. The particle dynamics is in the both cases determined by the matrix $\hat{L}$, i.e. on the grid with lattice constant $\Delta_x$. Obviously, the both approaches yield the same effective diffusion coefficient in the long-time limit. For the numerical calculation of the effective diffusion coefficient plotted in Fig.~\ref{fig:D}, we have used $i = N_x+1$ and $i + 1 = 0$. This procedure turned out to be much more stable even for a very fine discretization, we have used $N_x=N_y=1500$.

\section*{References}
\bibliographystyle{unsrt}

\end{document}